\documentclass[11pt]{article}
\usepackage{epsf}
\usepackage{color}
\begin{document}

\def\slash#1{{\rlap{$#1$} \thinspace/}}

\begin{flushright} 
September, 2004  \\
 YITP-04-51 \\
 KUNS-1934 \\
\end{flushright} 

\vspace{0.1cm}

\begin{Large}
\vspace{1cm}
\begin{center}
{\bf Fuzzy Supersphere and Supermonopole}
\end{center}
\end{Large}

\vspace{1cm}

\begin{center}
{\large Kazuki Hasebe $^{1}$ and Yusuke Kimura $^{2}$}   \\ 

\vspace{0.5cm} 
$^{1}$
Yukawa Institute for Theoretical Physics, Kyoto University, 
Kyoto 606-8502, Japan \\ 

$^{2}$
Department of Physics, Kyoto University, 
Kyoto 606-8502, Japan \\

\vspace{0.5cm} 
{\sf
hasebe@yukawa.kyoto-u.ac.jp} 

{\sf ykimura@gauge.scphys.kyoto-u.ac.jp}
\vspace{0.8cm} 

\end{center}

\vspace{1cm}

\begin{abstract}
\noindent
\end{abstract}
\hspace{0.4cm}
It is well-known that coordinates of a charged particle 
in a monopole background 
become noncommutative. 
In this paper, we study the motion of a charged particle 
moving on a supersphere in the presence of a supermonopole. 
We construct a supermonopole by using 
a supersymmetric extension of the first Hopf map. 
We investigate algebras of angular momentum operators 
and supersymmetry generators. 
It is shown that coordinates of the particle 
are described by fuzzy supersphere in the lowest Landau level. 
We find that 
there exist two kinds of degenerate wavefunctions 
due to the supersymmetry. 
Ground state wavefunctions are given by the Hopf spinor 
and we discuss their several properties.

\newpage 

\section{Introduction}
\hspace{0.4cm}
Over the past few years 
several papers have been devoted to the study of 
a relationship between noncommutative geometry and string theory. 
The need of noncommutative geometry in string theory is easily 
understood by considering a world-volume action of D-branes. 
D-branes are defined as the endpoints of open strings. 
Since gauge fields appear in the ground state of 
open strings, the low energy dynamics of D-branes is described 
by gauge fields. 
One of the most interesting aspects is the appearance 
of nonabelian gauge symmetry from the world-volume 
theory of some coincident D-branes, and transverse coordinates 
of $N$ D-branes are expressed by $U(N)$ adjoint scalars. 
The appearance of the matrix-valued coordinates 
implies a relationship between string theory and 
noncommutative geometry. 

The appearance of noncommutative geometry in string theory 
can be understood from a different point of view. 
It is also observed that a world volume theory on a D-brane 
in the presence of NS-NS two form background is described by 
noncommutative Yang-Mills theory \cite{hep-th/9908142}. 
We can say that noncommutative geometry appears 
in two different situations. 
A D$2$-brane can be constructed from 
multiple D$0$-branes by imposing a noncommutative relation 
on their coordinates. 
The size of matrix represents the number of 
D$0$-branes. 
On the other hand, world volume coordinates of 
a D2-brane under the strong magnetic field 
become noncommutative. The magnetic charge is interpreted 
as the number of D$0$-branes. 
These two descriptions are supposed to be same. 
As these examples show, to study these two descriptions 
leads to understanding a relationship 
between D-branes with different dimensions. 

The existence of these descriptions is easily understood 
by considering the quantum Hall system. 
It is well-known that 
noncommutative coordinates can be understood as 
guiding center coordinates in a strong magnetic field. 
The above two descriptions of D-branes are 
related to 
the existence of two kinds of coordinate, 
usual commutative coordinates and noncommutative guiding 
center coordinates. 
The appearance of noncommutative geometry 
in both theories is a common feature. 
By taking the lowest Landau limit or 
the zero slope limit (discussed in \cite{hep-th/9908142}), 
both theories obtain effective descriptions in terms of 
noncommutative geometry. 
A proposal given in \cite{hep-th/0101029} 
manifests the fact that 
the quantum Hall system is described by string theories. 

Another recent development 
in string theory is understanding of noncommutative superspace. 
If we consider string theories 
in the R-R field strength or graviphoton background, 
coordinates of superspace become non(anti)commutative 
\cite{hep-th/0302109,hep-th/0302078,hep-th/0305248}. 
Various aspects of noncommutative superspace have been 
studied in \cite{hep-th/0002084,hep-th/0104190,hep-th/0306215,
hep-th/0306226,
hep-th/0306237,hep-th/0306251,
hep-th/0307039,hep-th/0307076,hep-th/0307236,
hep-th/0308012,hep-th/0308021,
hep-th/0309076,hep-th/0401147,
hep-th/0402137,hep-th/0403290,
hep-th/0405278}. 
Some studies 
from the viewpoint of supermatrix models are 
found in \cite{hep-th/0303210,hep-th/0307060,hep-th/0311005}. 

As in the bosonic noncommutative geometry, it is important 
to investigate two descriptions of noncommutative superspace. 
In this paper, 
we consider the motion of a charged particle on a supersphere 
in a supermonopole background 
as a supersymmetric generalization of the quantum Hall system. 
We show a relationship between commutative coordinates 
and noncommutative guiding center coordinates. 
A noncommutative version of supersphere 
called fuzzy supersphere 
has been investigated 
in \cite{hep-th/9507074,math-ph/9804013,hep-th/0204170}. 
We expect that such a noncommutative space arises 
in the lowest Landau level. 
The reason for dealing with a (fuzzy) sphere is that 
the quantity such as the charge of D$0$-branes 
is given by a finite quantity. 
A noncommutative sphere is usually obtained by 
introducing a cut-off parameter for the angular momentum 
in a usual sphere. 
It is introduced as a monopole charge in the context 
of the quantum Hall system. 
The cut-off parameter is related to the number of D$0$-branes (quanta); 
therefore it can be finite for compact spaces. 
This is an advantage in order to compare two descriptions. 
The realization of noncommutative superspace in the lowest Landau level 
has also been reported in \cite{hep-th/0306251,hep-th/0311159}. 

The organization of this paper is as follows. 
We first review the (bosonic) two-sphere system 
in section \ref{sec:reviewofbosonictwospheresystem}. 
The Dirac monopole is 
introduced by the first Hopf map. 
According to the Hopf map, the gauge field is obtained 
from the so-called Hopf spinor. 
The Hopf spinor plays an important role 
in the quantum Hall system 
since it becomes a ground state eigenfunction 
of the Hamiltonian. 
We explain how a noncommutative space arises 
after we take the strong magnetic field limit. 
In section \ref{sec:supermonopoleandsupersphere}, 
we introduce a supersymmetric generalization of the Dirac monopole 
by using a supersymmetric generalization of the first Hopf map. 
The construction of the supermonopole is based on 
the method given in \cite{math-ph/9907020}. 
We explicitly construct the Hopf spinors 
for an arbitrary monopole charge. 
In section section \ref{sec:fuzzysphereinlll}, 
we analyze the motion of a particle moving on $S^{2,2}$. 
Symmetries of $S^{2,2}$ are given by Lie supergroup $OSp(1|2)$. 
The Hamiltonian of a free particle 
is written down in terms of the $osp(1|2)$ (and $osp(2|2)$)
generators. 
The contribution of the monopole 
is added by replacing usual derivatives 
with gauge covariant derivatives. 
The $osp(1|2)$ generators in the monopole background 
are deformed compared to those without the monopole background. 
We can obtain guiding center coordinates from 
the deformed $osp(1|2)$ generators. 
It is shown that commutative coordinates of a particle 
are identified with noncommutative guiding center coordinates 
in the lowest Landau level. They are found to satisfy the algebra of 
the fuzzy supersphere. 
Ground state wavefunctions are obtained from 
the Hopf spinors. We have two kinds of 
wavefunctions with the same energy 
because of the supersymmetry. We discuss their probability density and 
transformation property under the supersymmetry. 
Section \ref{sec:summary} 
is devoted to summary and discussions. 
Notations related to the superalgebra 
are summarized in appendix \ref{sec:notation}. 
In appendix \ref{sec:osp(2|2)algebra}, we comment on 
the $osp(2|2)$ algebra. The $osp(2|2)$ generators are constructed from 
the $osp(1|2)$ generators and play an important role in 
constructing the Hamiltonian. 
We show how they are deformed 
in the presence of the supermonopole. 
The representation theory of $OSp(1|2)$ and $OSp(2|2)$ is 
reviewed in appendix \ref{sec:repofosp22}. 
The detailed calculation of (\ref{commutationrelationofosp12lambda}) 
is presented in appendix \ref{sec:derivationof}. 

\section{Review of two-sphere system}
\label{sec:reviewofbosonictwospheresystem}
\hspace{0.4cm}
In this section, we review a (bosonic) two-sphere system. 
We consider a particle moving on a two-sphere in the background 
of a monopole put at the origin. 

Let us first introduce the Dirac monopole based on the first Hopf map.  
The first Hopf map is defined as a map from $S^{3}$ to $S^{2}$ 
which is expressed as 
\begin{equation}
x_{i}=2r\phi^{\dagger}\sigma_{\left(1/2\right)i}\phi, 
\label{hopfmap}
\end{equation}
where $\sigma_{\left(1/2\right)i}$ is 
the spin $1/2$ representation of $su(2)$ 
\footnote{
It is related to the Pauli matrix as 
$2\sigma_{\left(1/2\right)i}=\sigma_{i}$
}
. 
$\phi$ is a complex two-components spinor satisfying 
$\phi^{\dagger}\phi=1$ and is called Hopf spinor. 
$\phi^{\dagger}$ means the hermitian conjugate of $\phi$. 
The condition $\phi^{\dagger}\phi=1$ leads to 
$x_{i}x_{i}=r^{2}$. 
The Hopf spinor satisfying (\ref{hopfmap}) is explicitly given by 
\begin{equation}
\phi=
 \left( \begin{array}{c} 
 \phi_{1} \\ \phi_{2}
 \end{array} \right)=
 \frac{1}{\sqrt{2r(r+x_{3})}}
 \left( \begin{array}{c}
  r+x_{3}  \\ 
 x_{1}+ix_{2}    \\
 \end{array} \right)e^{i\chi}, 
 \label{hopfspinor}
\end{equation}
where $e^{i\chi}$ is a $U(1)$ phase. 
A $U(1)$ gauge transformation is generated 
by $\chi\rightarrow \chi+\Lambda$. 
A $U(1)$ gauge field is obtained from the Hopf spinor as 
\begin{equation}
A_{i}dx_{i}=-i\frac{\hbar}{e}\phi^{\dagger}d\phi 
=-\frac{g}{r(r+x_{3})}
\epsilon_{ij3}x_{j}dx_{i},
\end{equation} 
where $g \equiv \hbar/2e$ is the monopole charge. 
A monopole with $g=\hbar S/e$ is obtained by replacing 
$\phi$ with the following $(2S+1)$-components spinor: 
\begin{equation}
\phi_{(S,m)}=\sqrt{\frac{(2S)!}{(S-m)!(S+m)!}}
\phi_{1}^{S+m}\phi_{2}^{S-m}, 
\label{spinShopfspinor}
\end{equation}
where $2S$ is a positive integer, and 
$m$ takes values $-S, -S+1, \cdots,S$. 
The $S=1/2$ case corresponds to (\ref{hopfspinor}). 
The normalization is determined from the following condition,  
\begin{equation}
\int_{S^{2}} d \Omega \phi^{\ast}_{(S,m)}\phi_{(S,m^{\prime})}
=\frac{4\pi}{2S+1}\delta_{m,m^{\prime}}. 
\end{equation}
The equation (\ref{hopfmap}) is replaced with 
\begin{equation}
x_{i}=\frac{1}{S}r\phi_{(S)}^{\dagger}\sigma_{(S)i}\phi_{(S)}, 
\end{equation}
where $\sigma_{(S)i}$ is the spin $S$ representation of $su(2)$. 
This $x_{i}$ also satisfies $x_{i}x_{i}=r^{2}$. 
We note that 
this construction naturally realizes 
the Dirac quantization condition: 
\begin{equation}
eg=\hbar S. 
\label{Diraccondition}
\end{equation} 
The field strength of this monopole is
\begin{equation}
F_{ij}=\partial_{i}A_{j}-\partial_{j}A_{i}=
\frac{g}{r^{3}}\epsilon_{ijk}x_{k}. 
\end{equation}
The first Chern number is calculated as 
\begin{equation}
c_{1}=\frac{e}{2\pi \hbar }\int_{S^{2}} F=2S.
\label{firstChernbosonic}
\end{equation}

We next investigate the motion of a charged particle moving 
on a two-sphere in the monopole background. 
The Hamiltonian of such a particle is given by 
\begin{equation}
H=\frac{1}{2mr^{2}}\Lambda_{i}\Lambda_{i},
\end{equation}
where $m$ is the mass of the particle, and 
$\Lambda_{i}$ is the orbital 
angular momentum of the charged particle in the monopole 
background: 
\begin{equation}
\Lambda_{i}=
\epsilon_{ijk}x_{j}(-i\hbar\partial+eA)_{k}. 
\end{equation}
These 
$\Lambda_{i}$ no longer satisfy the algebra of the usual 
angular momentum and are deformed to 
\begin{equation}
[\Lambda_{i},\Lambda_{j}]=i\hbar\epsilon_{ijk}
\left(\Lambda_{k}-\frac{eg}{r} x_{k}\right). 
\end{equation}
Operators generating the  
$SU(2)$ rotation in the presence of the monopole are 
found to be 
\begin{equation}
L_{i}=\Lambda_{i}+\frac{eg}{r}x_{i}. 
\label{Llambdax}
\end{equation}
The last term represents the contribution from the monopole 
background, and 
$L_{i}$ can be interpreted to be the total angular momentum. 
They actually satisfy 
\begin{equation}
[L_{i},L_{j}]=i\hbar\epsilon_{ijk}L_{k}, \hspace{0.4cm}
[L_{i},\Lambda_{j}]=i\hbar\epsilon_{ijk}\Lambda_{k},
\hspace{0.4cm}
[L_{i},x_{j}]=i\hbar\epsilon_{ijk}x_{k}.
\end{equation}
From these relations, it is easily shown that 
$[L_{i},H]=0$, which manifests the fact that 
this system has the $SU(2)$ symmetry generated by $L_{i}$. 
We suppose the representation of $L_{i}$ to be the spin $l$. 
Then by using the following relation 
\begin{equation}
\Lambda_{i}\Lambda_{i}=
L_{i}L_{i}-(eg)^{2}
=\hbar^{2}\left(l(l+1)-S^{2}\right), 
\label{lambda^2}
\end{equation}
we can get the following energy eigenvalue of the Hamiltonian, 
\begin{equation}
E_{n}
=\frac{\hbar^{2}}{2mr^{2}}
\left( n(n+1)+(2n+1)S\right), 
\label{energyeigenvalue}
\end{equation}
where we have set $l=n+S \hspace{0.1cm}(n=0,1,2,\cdots)$. 
$n$ plays the role of the Landau level index 
and $n=0$ corresponds to the lowest Landau level. 
Since an energy interval between the lowest Landau level and 
the first Landau level is given by $\Delta E=S\hbar^{2}/mr^{2}$, 
the motion of the particle is confined to the lowest Landau level 
in the strong magnetic field limit: 
\begin{equation}
S/mr \gg 1. 
\label{strongmagneticfieldlimit}
\end{equation}
The degeneracy of the lowest Landau level is $2S+1$. 
It is related to the size of noncommutative 
space as we will see later. 

As in the well-known planar system, 
the motion of the charged particle obeys the cyclotron motion. 
The guiding center coordinates $X_{i}$ can be introduced as 
\begin{equation}
X_{i} =\alpha L_{i}, \hspace{0.4cm} 
\alpha \equiv \frac{r}{eg}
.
\end{equation}
They satisfy the following noncommutative relation 
\begin{equation}
[X_{i},X_{j}]=i\hbar\alpha\epsilon_{ijk} X_{k}. 
\label{noncommutativesphere}
\end{equation}
From the equation (\ref{Llambdax}), we obtain 
a relationship between the guiding center coordinates 
and the commutative coordinates as
\begin{equation}
X_{i}=\alpha\Lambda_{i}+x_{i}. 
\label{guidingcenterrelation}
\end{equation}
The radius of the cyclotron motion 
in the $n$-th Landau level is evaluated as 
\begin{equation}
r_{n}^{cyc}=\alpha \hbar \sqrt{n(n+1)+(2n+1)S}. 
\end{equation}
In the lowest Landau level, it becomes 
\begin{equation}
r_{0}^{cyc}=\alpha \hbar\sqrt{S}=\frac{r}{\sqrt{S}}. 
\end{equation}
Since the radius $r_{0}^{cyc}$ 
becomes much smaller than $r$ 
in the strong magnetic field limit 
(\ref{strongmagneticfieldlimit}), 
the commutative coordinates $x_{i}$ are identified 
with the noncommutative coordinates $X_{i}$ 
in the lowest Landau level. 
The noncommutative geometry described by $X_{i}$ is known as 
fuzzy sphere. 
The radius of the cyclotron motion for the ground state 
provides the noncommutative length: 
$l_{NC}\equiv r_{0}^{cyc}$. 
The radius of the fuzzy sphere is given by the quadratic Casimir 
of $su(2)$ as 
\begin{equation}
r^{2}
=\hbar^{2}\alpha^{2}S(S+1).
\end{equation}
If we substitute $\alpha=r/eg$, 
the Dirac quantization condition (\ref{Diraccondition}) 
is reproduced in the large $S$ limit. 

We shall consider the thermodynamic limit. 
It is given by 
the large $S$ limit with keeping 
the noncommutative scale $l_{NC}$ finite. 
In this limit, the energy eigenvalue (\ref{energyeigenvalue}) 
approaches 
\begin{equation}
E_{n}\rightarrow \frac{\hbar^{2}}{m l_{NC}^{2}}
\left(n+\frac{1}{2}\right). 
\label{spectruminthermodynamic}
\end{equation}
This corresponds to the planar Landau levels. 

Before finishing this section, we comments on 
the eigenstates of this system. 
When $S=1/2$, the Hopf spinor (\ref{hopfspinor}) is found to 
become the ground state wavefunction of the Hamiltonian. 
In general, 
the eigenstate with the eigenvalue $E_{0}$ in 
(\ref{energyeigenvalue})
is given by the Hopf spinor (\ref{spinShopfspinor}). 
It is because the Hopf spinor (\ref{hopfspinor}) 
(and (\ref{spinShopfspinor}))
transforms as an $SU(2)$ spinor. 
We should notice that the conjugate spinor 
$\bar{\phi}$ does not enter the 
eigenstate in the lowest Landau level. 
This fact is an analogous to 
the result in the planar system 
where 
wavefunctions in the lowest Landau level are written 
in terms of polynomials of only $z$ 
(up to a Gaussian factor). 
The probability density of the eigenstates is 
given by 
\begin{eqnarray}
|\phi_{(S,m)}|^{2}=\frac{(2S)!}{(S-m)!(S+m)!}
\left(\frac{1}{2r}\right)^{2S}
\left(r+x_{3}\right)^{S+m}
\left(r-x_{3}\right)^{S-m}. \cr
\end{eqnarray}
This state forms a ring and is localized at $x_{3}=(m/S)r$. 
This result reminds us of 
the planar system in the symmetric gauge.


\section{Supermonopole}
\label{sec:supermonopoleandsupersphere}
\hspace{0.4cm}
In the previous section, we reviewed the bosonic two-sphere 
system and observed that coordinates of a charged particle 
are described by the fuzzy two-sphere 
in the lowest Landau level. In the following sections, 
we study the motion of a charged particle moving 
on the supersphere $S^{2,2}$ 
as a supersymmetric generalization of the previous section. 
\footnote{
A similar analysis for the superspace 
$SU(1|2)/[U(1)\times U(1)]$ has been made in \cite{hep-th/0311159}. 
} 
We expect that the coordinates are described by 
the fuzzy supersphere in the same way as the bosonic case. 

We first review the supersphere. 
The supersphere $S^{2,2}$ is characterized by the coset space 
given by $OSp(1|2)/U(1)$. 
Let $x_{i}$ ($i=1,2,3$) and 
$\theta_{\alpha}$ ($\alpha=1,2$) 
be coordinates of the supersphere which are related as 
\begin{equation}
x_{i}x_{i}+C_{\alpha\beta}\theta_{\alpha}\theta_{\beta}=r^{2}, 
\label{relationxandtheta} 
\end{equation}
where $C_{\alpha\beta}$ is the antisymmetric tensor with $C_{12}=1$. 
We define a coordinate $y_{i}$ such that $y_{i}y_{i}=r^{2}$. 
A space which is defined by the coordinate $y_{i}$ is 
called the {\it body} of the superspace. 
Hence $S^{2}$ is the body of $S^{2,2}$. 
It is related to $x_{i}$ as 
\begin{equation}
y_{i}=\left(1+\frac{\theta C\theta}{2r^{2}}\right)x_{i}.
\end{equation}
The remaining coordinate $\theta_{\alpha}$ is called the {\it soul}. 

The supersphere has an $SU(2)$ rotational symmetry and 
supersymmetry which are generated by 
\begin{eqnarray}
&&J_{i}=-i\hbar\epsilon_{ijk}x_{j}\partial_{k}
+\frac{1}{2}\hbar\theta_{\alpha}(\sigma_{i})_{\alpha\beta}
\partial_{\beta}, 
\cr
&&J_{\alpha}=\frac{1}{2}\hbar x_{i}
(C\sigma_{i})_{\alpha\beta}\partial_{\beta}
-\frac{1}{2}\hbar\theta_{\beta}(\sigma_{i})_{\beta\alpha}\partial_{i},
\label{osp12generator}
\end{eqnarray}
respectively. 
They satisfy the following $osp(1| 2)$ algebra, 
\begin{equation}
[J_{i},J_{j}]=i\hbar\epsilon_{ijk}J_{k}, 
\hspace{0.4cm}
[J_{i},J_{\alpha}]=\frac{1}{2}\hbar
(\sigma_{i})_{\beta\alpha}J_{\beta}, 
\hspace{0.4cm}
\{J_{\alpha},J_{\beta}\}
=\frac{1}{2}\hbar(C\sigma_{i})_{\alpha\beta}J_{i}.  
\end{equation}
The $osp(1| 2)$ algebra is simply reviewed 
in appendix \ref{sec:repofosp22}. 
The coordinates transform under the supersymmetry as  
\begin{eqnarray}
&&\delta x_{i}=\frac{1}{2}(\epsilon \sigma_{i}C\theta), \cr
&&\delta \theta_{\alpha}=-\frac{1}{2}
(\epsilon \sigma_{i})_{\alpha}x_{i} , 
\label{susy}
\end{eqnarray}
where $\epsilon_{\alpha}$ are Grassmann parameters. 
The radius of $S^{2,2}$ is invariant under the supersymmetry 
\begin{equation}
\delta r=0. 
\end{equation}

Let us next introduce a supersymmetric generalization 
of the Dirac monopole. 
We use a supersymmetric generalization of 
the first Hopf map $S^{3,2}\rightarrow S^{2,2}$ 
based on \cite{math-ph/9907020}. 
We will obtain an explicit form of the Hopf spinor 
expressed by the coordinate of $S^{2,2}$. 
It plays an important role since 
it becomes a wavefunction in the lowest Landau level 
as is discussed in the next section. 
The map is expressed by 
\begin{equation}
x_{i}=2r\phi^{\ddagger}l_{i} \phi, \hspace{0.4cm}
\theta_{\alpha}=2r \phi^{\ddagger} v_{\alpha} \phi, 
\label{superhopfmap}
\end{equation}
where 
\begin{equation}
l_{i}= \frac{1}{2}\left( \begin{array}{cc} 
 \sigma_{i}& 0\\ 
 0&0 
  \end{array}\right), \hspace{0.4cm}
  v_{1}=\frac{1}{2} \left( \begin{array}{cc} 
 0& \eta\\ 
 \xi^{T} &0 
  \end{array}\right),\hspace{0.4cm}
    v_{2}= \frac{1}{2}\left( \begin{array}{cc} 
 0& \xi\\ 
 -\eta^{T} &0 
  \end{array}\right)
\end{equation}
are the three dimensional representation of $osp(1|2)$, 
and 
\begin{equation}
\eta=\left( \begin{array}{c} 
 -1\\ 
 0 
  \end{array}\right), \hspace{0.4cm}
  \xi=\left( \begin{array}{c} 
 0\\ 
 -1 
  \end{array}\right).
\end{equation}
$\phi$ is a complex three-components spinor which satisfies 
\begin{equation}
  \phi^{\ddagger} \phi=1,
  \label{phiddaggerphi=1}
\end{equation}
where $\phi^{\ddagger}$ is defined as 
$(\phi_{1}^{\ast},\phi_{2}^{\ast},-\psi^{\ast})$. 
It must be noted that the minus sign is added to the third component.  
An explicit form of $\phi$ is given by 
the coordinates of $S^{2,2}$ as 
\begin{eqnarray}
&&\phi= \left( \begin{array}{c} 
 \phi_{1} \\ \phi_{2} \\ \psi
 \end{array} \right)
 =\frac{1}{\sqrt{2r^{3}(r+x_{3})}}
 \left( \begin{array}{c}
  (r+x_{3})
  \left(r-\frac{1}{4(r+x_{3})}\theta C\theta\right)  \\ 
 (x_{1}+ix_{2})
  \left(r+\frac{1}{4(r+x_{3})}\theta C\theta\right)  \\
   -
  \left((r+x_{3})\theta_{1}+(x_{1}+ix_{2})\theta_{2}\right) 
 \end{array} \right) e^{i\chi} \cr
 {\rule[-2mm]{0mm}{12mm}\ } 
&&\hspace{2.2cm} 
 =\frac{1}{\sqrt{2r^{3}(r+y_{3})}}
 \left( \begin{array}{c}
  (r+y_{3})
  \left(r-\frac{1}{4r}\theta C\theta\right)  \\ 
 (y_{1}+iy_{2})
  \left(r-\frac{1}{4r}\theta C\theta\right)  \\
  -\left((r+y_{3})\theta_{1}+(y_{1}+iy_{2})\theta_{2}\right) 
 \end{array} \right) e^{i\chi}
 \label{superhopfspinorS=1/2}
\end{eqnarray}
where $\chi$ is a bosonic coordinate and 
$e^{i\chi}$ is a $U(1)$ phase factor. 
A $U(1)$ local gauge transformation is induced by 
$\chi \rightarrow \chi+\Lambda$. 
From this explicit representation, the equation 
(\ref{phiddaggerphi=1}) is checked 
by making use of (\ref{probabolityofS=1/2}).
A $U(1)$ gauge field is obtained from the Hopf spinor $\phi$ as 
\begin{equation}
A_{i}dx_{i}+A_{\alpha}d\theta_{\alpha}
=-i\frac{\hbar}{e}\phi^{\ddagger}d \phi. 
\end{equation} 
Hence each component of $A$ is obtained as 
\begin{eqnarray}
&&A_{i}=-\frac{g}{r(r+x_{3})}\epsilon_{ij3}x_{j}
\left(1+\frac{2r+x_{3}}{2r^{2}(r+x_{3})}\theta C\theta\right) 
 \cr
 &&\hspace{0.51cm}
 =-\frac{g}{r(r+y_{3})}\epsilon_{ij3}y_{j}
\left(1+\frac{1}{2r^{2}}\theta C\theta\right), 
\cr
&&A_{\alpha}=\frac{ig}{r^{3}}x_{i}(\theta\sigma_{i}C)_{\alpha} 
\cr 
&&\hspace{0.5cm}
=\frac{ig}{r^{3}}y_{i}(\theta\sigma_{i}C)_{\alpha}, 
\label{super1/2gaugefield}
\end{eqnarray}
where $g \equiv \hbar/2e$ is the monopole charge. 
Note that $A$ satisfies the reality condition $A^{\ddagger}=A$. 
This gauge field is singular at the south pole. We can construct 
the gauge field which is singular at the north pole 
by using the following Hopf spinor: 
\begin{eqnarray}
&&\phi^{\prime}= \left( \begin{array}{c} 
 \phi_{1}^{\prime} \\ \phi_{2}^{\prime} \\ \psi^{\prime}
 \end{array} \right)
 =\frac{1}{\sqrt{2r^{3}(r-x_{3})}}
 \left( \begin{array}{c}
 (x_{1}-ix_{2})
  \left(r+\frac{1}{4(r-x_{3})}\theta C\theta\right)  \\
  (r-x_{3})
  \left(r-\frac{1}{4(r-x_{3})}\theta C\theta\right)  \\  
   -
  \left((x_{1}-ix_{2})\theta_{1}+(r-x_{3})\theta_{2}\right) 
 \end{array} \right) e^{i\chi} . 
\end{eqnarray}
The corresponding gauge field is 
\begin{eqnarray}
&&A_{i}^{\prime}=\frac{g}{r(r-x_{3})}\epsilon_{ij3}x_{j}
\left(1+\frac{2r-x_{3}}{2r^{2}(r-x_{3})}\theta C\theta\right), 
\cr
&&A_{\alpha}^{\prime}=\frac{ig}{r^{3}}x_{i}(\theta\sigma_{i}C)_{\alpha}. 
\label{super1/2gaugefield2}
\end{eqnarray}
(\ref{super1/2gaugefield}) and (\ref{super1/2gaugefield2}) 
are related by the gauge transformation such that 
the gauge parameter is given by 
$\tan \Lambda =x_{2}/x_{1}$. 
A monopole which has a larger charge is obtained by using 
the following Hopf spinor, 
\begin{eqnarray}
&&\Phi_{(S,m)}=\sqrt{\frac{(2S)!}{(S-m)!(S+m)!}}
\phi_{1}^{S+m}\phi_{2}^{S-m}, 
\cr
&&\Psi_{(S,m^{\prime})}=\sqrt{\frac{(2S)!}
{\left(S-1/2-m^{\prime}\right)!\left(S-1/2+m^{\prime}\right)!}}
\phi_{1}^{S-1/2+m^{\prime}}\phi_{2}^{S-1/2-m^{\prime}}\psi, 
\label{supergeneralhopfspinor}
\end{eqnarray}
where $m$ runs over $-S,-S+1,\cdots,S$, and 
$m^{\prime}$ over $-S+1/2,-S+3/2,\cdots,S-1/2$. 
The orthonormal relations are 
\begin{eqnarray}
&&\int_{S^{2,2}}d\Omega_{(2,2)}
\Phi_{(S,m)}^{\ast}\Phi_{(S,m^{\prime})}
=\frac{8\pi S}{2S+1} \delta_{m,m^{\prime}}\cr
&&\int_{S^{2,2}}d\Omega_{(2,2)}
\Psi_{(S,m)}^{\ast}\Psi_{(S,m^{\prime})}
=4\pi \delta_{m,m^{\prime}} \cr
&&\int_{S^{2,2}}d\Omega_{(2,2)}
\Phi_{(S,m)}^{\ast}\Psi_{(S,m^{\prime})}
=0 ,
\end{eqnarray}
where we have defined $d\Omega_{(2,2)}=d\Omega_{S^{2}} 
d\theta_{1}d\theta_{2}$. 
In this case, the relation (\ref{superhopfmap}) is 
modified to 
\begin{equation}
x_{i}=\frac{1}{S}r\phi_{(S)}^{\ddagger}l_{(S)i} \phi_{(S)}, 
\hspace{0.4cm}
\theta_{\alpha}=\frac{1}{S}r \phi_{(S)}^{\ddagger} v_{(S)\alpha}\phi_{(S)}, 
\end{equation}
where $\phi_{(S)}\equiv(\Phi_{(S)},\Psi_{(S)})^{T}$, and 
 $l_{(S)i}$ and $v_{(S)\alpha}$ are 
$(4S+1)$-dimensional representation of $osp(1|2)$. 
The gauge field strength is calculated as 
\begin{eqnarray}
&&F=dA=\frac{1}{2}F_{ij}dx_{i}\wedge dx_{j}
+F_{i\alpha}dx_{i}\wedge d\theta_{\alpha}
+\frac{1}{2}F_{\alpha\beta}
d\theta_{\alpha}\wedge d\theta_{\beta} \cr
&&\hspace{1.4cm}
=\frac{1}{2}(\partial_{i}A_{j}-\partial_{j}A_{i})
dx_{i}\wedge dx_{j}+
(\partial_{i}A_{\alpha}-\partial_{\alpha}^{R}A_{i})
dx_{i}\wedge d\theta_{\alpha} \cr
&&\hspace{1.8cm}
+\frac{1}{2}(\partial_{\alpha}A_{\beta}+\partial_{\beta}A_{\alpha})
d\theta_{\alpha}\wedge d\theta_{\beta},
\end{eqnarray}
where we have used the notation such as 
$\partial_{\alpha}A=(\partial /\partial\theta_{\alpha})A$ and 
$\partial_{\alpha}^{R}A=\partial A /\partial\theta_{\alpha}$. 
Hence we get 
\begin{eqnarray}
&&F_{ij}=\frac{g}{r^{3}}\epsilon_{ijk}x_{k}
\left(1+\frac{3}{2r^{2}}\theta C\theta \right) \cr
&&\hspace{0.6cm}
=\frac{g}{r^{3}}\epsilon_{ijk}y_{k}
\left(1+\frac{1}{r^{2}}\theta C\theta \right), \cr
{\rule[-2mm]{0mm}{10mm}\ } 
&&F_{i\alpha}=-\frac{2g}{r^{3}}i
\left(\frac{3}{2}\frac{x_{i}x_{j}}{r^{2}}-\frac{1}{2}\delta_{ij}\right)
(\theta\sigma_{j}C)_{\alpha} \cr
&&\hspace{0.6cm}
=-\frac{2g}{r^{3}}i
\left(\frac{3}{2}\frac{y_{i}y_{j}}{r^{2}}-\frac{1}{2}\delta_{ij}\right)
(\theta\sigma_{j}C)_{\alpha} ,\cr
{\rule[-2mm]{0mm}{10mm}\ } 
&&F_{\alpha\beta}=
i\frac{2g }{r^{3}}x_{i}(\sigma_{i}C)_{\alpha\beta}
\left(1+\frac{3}{2r^{2}}\theta C\theta \right) \cr
&&\hspace{0.7cm}
=i\frac{2g }{r^{3}}y_{i}(\sigma_{i}C)_{\alpha\beta}
\left(1+\frac{1}{r^{2}}\theta C\theta \right),
\label{chargeSmonopole}
\end{eqnarray}
where the monopole charge is $g=\hbar S/e$. 

We next see how the above components transform under 
the supersymmetry (\ref{susy}). 
Defining the bosonic part of the magnetic field as 
\begin{equation}
B_{i}\equiv \frac{g}{r^{3}}x_{i}\left(1+\frac{3}{2r^{2}}
\theta C\theta \right), 
\end{equation}
we obtain 
\begin{eqnarray}
&&\delta B_{i}=-i\epsilon_{\alpha} F_{i\alpha}, \cr
&&\delta F_{i\alpha}
=-\frac{1}{4}\epsilon_{ijk}(\epsilon \sigma_{k}C)_{\alpha}
B_{j}+\frac{i}{2}(\epsilon C)_{\alpha}B_{i}. 
\label{susyfieldstrength}
\end{eqnarray}
We can recognize that $B_{i}$ and $F_{i\alpha}$ form 
a multiplet under the supersymmetry (\ref{susy}). 

Let us calculate the first Chern character of the supermonopole 
\cite{math-ph/9907020}. 
We define it as 
\begin{eqnarray}
c_{1}&=&\frac{e}{2\pi \hbar} \int_{S^{2,2}} F \cr 
&=& \frac{e}{2\pi \hbar} \int_{S^{2,2}}\left(\frac{1}{2}
F_{ij}dx_{i}\wedge dx_{j}
+F_{i\alpha}dx_{i}\wedge d\theta_{\alpha}
+\frac{1}{2}F_{\alpha\beta}d\theta_{\alpha}\wedge d\theta_{\beta} 
\right). 
\label{firstChernofsupermonopole}
\end{eqnarray}
The important point is that 
the coordinate $x_{i}$ depends on the Grassmann coordinates 
due to the relation (\ref{relationxandtheta}):
\begin{equation}
\sqrt{x_{i}x_{i}}=\sqrt{r^{2}-\theta C\theta}
=r-\frac{1}{2r}\theta C\theta. 
\label{grassmanninxx}
\end{equation}
The integration over the Grassmann variables is evaluated by 
the Berezin integral. 
It is found that 
the second and third terms in (\ref{firstChernofsupermonopole}) 
vanish by integrating the Grassmann coordinates. 
As for the first term, the dependence of the Grassmann coordinates 
in $F_{ij}$ cancels by that in $dx_{i}\wedge dx_{j}$ 
which comes from (\ref{grassmanninxx}). 
Consequently the integral over the supersphere 
results in the integral over the body: 
\begin{equation}
c_{1}
=\frac{e}{2\pi \hbar}
\int_{S^{2}} \frac{1}{2}
F_{ij\mid \theta=0}dy_{i}\wedge dy_{j} =2S. 
\label{superchernnumber}
\end{equation}
We have obtained the same result 
as the bosonic case (\ref{firstChernbosonic}).


\section{Fuzzy supersphere as the lowest Landau level}
\label{sec:fuzzysphereinlll}
\hspace{0.4cm}
In this section, we analyze the motion of a particle 
moving on $S^{2,2}$ in the presence of the supermonopole 
background 
and see how noncommutative superspace arises in the lowest 
Landau level. 

The Hamiltonian we start with is the following 
\footnote{This Hamiltonian does not 
provide a complete form of the kinetic term 
of a particle moving on $S^{2,2}$ 
though it is a supersymmetric generalization of 
the bosonic case. 
We, nevertheless, use this Hamiltonian 
since it is the simplest supersymmetric generalization and 
enables us to 
know some properties peculiar to supersymmetric systems. 
The {\it correct} Hamiltonian 
is given in appendix \ref{sec:osp(2|2)algebra}. }, 
\begin{equation}
H=\frac{1}{2mr^{2}}\left(\Lambda_{i}\Lambda_{i}
+C_{\alpha\beta}\Lambda_{\alpha}\Lambda_{\beta}\right). 
\end{equation}
$\Lambda_{i}$ and $\Lambda_{\alpha}$ are the gauge 
covariant operators which are obtained 
from (\ref{osp12generator}) 
by making the following replacements, 
\begin{eqnarray}
&&\partial_{i}\rightarrow \partial_{i}+ieA_{i}, \cr 
&&\partial_{\alpha}\rightarrow \partial_{\alpha}-ieA_{\alpha}. 
\label{gaugefieldreplacement}
\end{eqnarray}
$(\Lambda_{i},\Lambda_{\alpha})$ are orthogonal to the coordinates 
$(x_{i},\theta_{\alpha})$: 
\begin{equation}
x_{i}\Lambda_{i}+\theta C \Lambda=0. 
\label{orthogonalrelation}
\end{equation}
Since we have replaced the derivative 
with the gauge covariant derivatives, 
$\Lambda_{i}$ and $\Lambda_{\alpha}$ no longer satisfy the 
$osp(1|2)$ algebra. Their commutation relations 
become 
\begin{eqnarray}
&&[\Lambda_{i},\Lambda_{j}]=i\hbar\epsilon_{ijk}
\left(\Lambda_{k}-\frac{eg}{r}x_{k}\right), \cr 
{\rule[-2mm]{0mm}{5mm}\ } 
&&[\Lambda_{i},\Lambda_{\alpha}]=
\frac{1}{2}\hbar(\sigma_{i})_{\beta\alpha}
\left(\Lambda_{\beta}-\frac{eg}{r}\theta_{\beta}\right), \cr
{\rule[-2mm]{0mm}{5mm}\ } 
&&\{\Lambda_{\alpha},\Lambda_{\beta}\}=
\frac{1}{2}\hbar(C\sigma_{i})_{\alpha\beta}
\left(\Lambda_{i}-\frac{eg}{r}x_{i}\right). 
\label{commutationrelationofosp12lambda}
\end{eqnarray}
The detailed derivation of these relations 
is shown in appendix \ref{sec:derivationof}. 
Therefore, 
the $osp(1|2)$ generators 
in the supermonopole background are given by 
\begin{eqnarray}
&&L_{i}\equiv\Lambda_{i}+\frac{1}{\alpha}x_{i}, \cr 
{\rule[-2mm]{0mm}{6mm}\ } 
&&L_{\alpha}\equiv\Lambda_{\alpha}+\frac{1}{\alpha}\theta_{\alpha}, 
\label{LandLambda}
\end{eqnarray}
where $\alpha\equiv r/eg=r/\hbar S$. 
They satisfy 
\begin{equation}
[L_{i},L_{j}]=i \hbar\epsilon_{ijk}L_{k}, \hspace{0.4cm}
[L_{i},L_{\alpha}]=
\frac{1}{2}\hbar(\sigma_{i})_{\beta\alpha}L_{\beta}, \hspace{0.4cm}
\{L_{\alpha},L_{\beta}\}=
\frac{1}{2}\hbar(C\sigma_{i})_{\alpha\beta}L_{i}, 
\end{equation}
and 
\begin{eqnarray}
&&[L_{i},\Lambda_{j}]=i\hbar \epsilon_{ijk}\Lambda_{k}, 
\hspace{0.4cm}
[L_{i},x_{j}]=i\hbar\epsilon_{ijk}x_{k}, \cr 
&&[L_{i},\Lambda_{\alpha}]=
\frac{1}{2}\hbar(\sigma_{i})_{\beta\alpha}\Lambda_{\beta}, 
\hspace{0.4cm}
[L_{i},\theta_{\alpha}]=
\frac{1}{2}\hbar(\sigma_{i})_{\beta\alpha}\theta_{\beta} \cr 
&&[L_{\alpha},\Lambda_{i}]=-
\frac{1}{2}\hbar(\sigma_{i})_{\beta\alpha}\Lambda_{\beta}, 
\hspace{0.4cm}
[L_{\alpha},x_{i}]=-
\frac{1}{2}\hbar(\sigma_{i})_{\beta\alpha}\theta_{\beta}, \cr 
&&\{L_{\alpha},\Lambda_{\beta}\}=
\frac{1}{2}\hbar(C\sigma_{i})_{\alpha\beta}\Lambda_{i}, 
\hspace{0.4cm}
\{L_{\alpha},\theta_{\beta}\}=
\frac{1}{2}\hbar(C\sigma_{i})_{\alpha\beta}x_{i}. 
\end{eqnarray}
They also satisfy 
\begin{equation}
x_{i}L_{i}+\theta C L=\hbar rS. 
\label{orthogonalrelation2}
\end{equation}

Let us now suppose that $L_{i}$ and 
$L_{\alpha}$ belong to the superspin $l$ representation 
of $OSp(1|2)$ whose dimension is $N=4l+1$. 
The quadratic Casimir is given by 
\begin{equation}
L_{i}L_{i}+C_{\alpha\beta}L_{\alpha}L_{\beta}
=\hbar^{2}l\left(l+\frac{1}{2}\right). 
\end{equation}
We then have 
\begin{eqnarray}
\Lambda_{i}\Lambda_{i}+
C_{\alpha\beta}\Lambda_{\alpha}\Lambda_{\beta} 
&=&L_{i}L_{i}+
C_{\alpha\beta}L_{\alpha}L_{\beta}-\hbar^{2}S^{2} \cr
&=&\hbar^{2}\left(
l\left(l+\frac{1}{2}\right)-S^{2}
\right), 
\end{eqnarray}
where we have used the equation (\ref{orthogonalrelation2}). 
Using this equation, 
the energy eigenvalue of the Hamiltonian is found to be 
\begin{equation}
E_{n}= \frac{\hbar^{2}}{2mr^{2}}\left(
n\left(n+\frac{1}{2}\right)+
\left(2n+\frac{1}{2}\right)S \right), 
\label{superenergyeigenvalue}
\end{equation}
where we have set $l=n+S$ ($n=0,1,\cdots$). 
The integer $n$ characterizes the Landau level. 
It can be shown that the Hamiltonian has 
the $osp(1|2)$ symmetry
\begin{equation}
[L_{i},H]=[L_{\alpha},H]=0. 
\end{equation}
This means that there exist a degeneracy generated by 
$L_{i}$ and $L_{\alpha}$, 
which 
is related to the extension of a noncommutative superspace 
realized in the lowest Landau level 
(as will be seen later). 

We define the guiding center coordinates as 
\begin{eqnarray}
&& X_{i}=\alpha L_{i}=\alpha \Lambda_{i}+x_{i}, \cr 
&& \Theta_{\alpha}=\alpha L_{\alpha}
=\alpha\Lambda_{\alpha}+\theta_{\alpha}.
\end{eqnarray}
Noncommutative geometry is obtained in the similar way 
to the bosonic system. 
The motion of the particle is confined to the lowest Landau level 
by taking the large $S$ limit (\ref{strongmagneticfieldlimit}). 
The radius of the cyclotron motion 
in the $n$-th Landau level is now given by 
\begin{equation}
r_{n}^{scyc}=\alpha \hbar \sqrt{n(n+1/2)+(2n+1/2)S}. 
\end{equation}
The radius in the ground state ($n=0$) becomes much smaller 
than the radius of the supersphere $r$ in the large $S$ limit; 
accordingly 
the coordinates ($x_{i}$,$\theta_{\alpha}$) 
are identified with the noncommutative guiding center coordinates 
($X_{i}$,$\Theta_{\alpha}$). 
The coordinates are given by 
the superspin $S$ representation of $OSp(1|2)$ 
and form the following algebra, 
\begin{eqnarray}
&&[X_{i},X_{j}]=i\alpha\hbar\epsilon_{ijk}X_{k}, \cr
&&[X_{i},\Theta_{\alpha}]=\frac{1}{2}\alpha \hbar
(\sigma_{i})_{\beta\alpha}\Theta_{\beta}, \cr 
&&\{\Theta_{\alpha},\Theta_{\beta}\}
=\frac{1}{2}\alpha\hbar(C\sigma_{i})_{\alpha\beta}X_{i}. 
\end{eqnarray}
The superspin $S$ representation of 
$OSp(1|2)$ is given by a $(4S+1)\times (4S+1)$ matrix 
and is decomposed into the spin $S$ and $(S-1/2)$ 
representation of $SU(2)$. 
$L_{i}$ and $L_{\alpha}$ generate 
the $SU(2)$ rotation and supersymmetry respectively, acting on 
the noncommutative coordinates as 
\begin{eqnarray}
&&[L_{i},X_{j}]=i\hbar\epsilon_{ijk}X_{k}, \hspace{0.4cm}
[L_{i},\Theta_{\alpha}]=\frac{1}{2}\hbar
(\sigma_{i})_{\beta\alpha}\Theta_{\beta}, \cr 
&&[L_{\alpha},X_{i}]=-\frac{1}{2}\hbar
(\sigma_{i})_{\beta\alpha}\Theta_{\beta}, \hspace{0.4cm}
\{L_{\alpha},\Theta_{\beta}\}
=\frac{1}{2}\hbar(C\sigma_{i})_{\alpha\beta}X_{i}. 
\end{eqnarray}
The radius of the fuzzy supersphere is provided by 
\begin{equation}
r^{2}=X_{i}X_{i}+C_{\alpha\beta}\Theta_{\alpha}\Theta_{\beta}
=\alpha^{2}\hbar^{2}S\left(S+\frac{1}{2}\right). 
\end{equation}
The thermodynamic limit is given by the large $S$ limit 
with keeping noncommutative scale $l_{NC}$ finite. 
In this limit, (\ref{superenergyeigenvalue}) becomes 
\begin{equation}
E_{n}\rightarrow \frac{\hbar^{2}}{m l_{NC}^{2}}
\left(n+\frac{1}{4}\right). 
\end{equation}
We find that the ground state energy is lower than 
that of the bosonic system (\ref{spectruminthermodynamic}). 
This would be explained by the supersymmetry. 


We discuss the eigenfunctions in the lowest Landau level. 
The Hopf spinor 
(\ref{supergeneralhopfspinor}) becomes 
the eigenfunctions in the lowest Landau level 
since it is an $OSp(1|2)$ spinor. 
We also note that conjugate spinors do not appear 
in their expressions. 
A novel aspect compared to the bosonic system is 
the existence of the supersymmetry. 
Hence we have two 
kinds of eigenstates with the same energy. 
We can explicitly confirm that they are related by 
the supersymmetry transformation. 
For the superspin $1/2$ states, we have 
\begin{equation}
\left( \begin{array}{c} 
 \delta\Phi \\ \delta\Psi
 \end{array} \right)=
 \left( \begin{array}{c} 
 \delta\phi_{1} \\ \delta\phi_{2} \\ \delta\psi
 \end{array} \right)=
 \frac{1}{2}
 \left( \begin{array}{c}
  -\epsilon_{2} \psi \\ 
  \epsilon_{1} \psi  \\
  \epsilon_{2}\phi_{2}+\epsilon_{1}\phi_{1}
 \end{array} \right), 
\end{equation}
where $\epsilon_{\alpha}$ ($\alpha=1,2$) are Grassmann parameters. 
The probability density of these states is calculated as 
\begin{eqnarray}
&&|\phi_{1}|^{2}\equiv \phi_{1}^{\ast}\phi_{1}=\frac{r+y_{3}}{2r}
\left(1-\frac{1}{2r^{2}}\theta C\theta\right) ,\cr
&&|\phi_{2}|^{2}\equiv \phi_{2}^{\ast}\phi_{2}=\frac{r-y_{3}}{2r}
\left(1-\frac{1}{2r^{2}}\theta C\theta\right) ,\cr
&&|\psi|^{2}\equiv -\psi^{\ast}\psi=
\frac{1}{2r^{2}} \theta C\theta.  
\label{probabolityofS=1/2}
\end{eqnarray}
The first (second) state is localized at the north (south) pole, 
$y_{3}=r$ ($-r$). 
The third one does not depend on the coordinates of the body. 
These results can be generalized to the case of 
the superspin $S$. 
The supersymmetry transformation is given by 
\begin{eqnarray}
&&\delta \Phi_{(S,m)}
=\frac{1}{2}\left(-\epsilon_{2}\sqrt{S+m}\Psi_{(S,m-1/2)}
+\epsilon_{1}\sqrt{S-m}\Psi_{(S,m+1/2)}\right), \cr
&&\delta \Psi_{(S,m^{\prime})}
=\frac{1}{2}\left(\epsilon_{2}\sqrt{S+1/2-m^{\prime}}
\Phi_{(S,m^{\prime}-1/2)}
+\epsilon_{1}\sqrt{S+1/2+m^{\prime}}
\Phi_{(S,m^{\prime}+1/2)}\right) . 
\label{superspinSwavefnctionsusytr}
\end{eqnarray}
The probability density for these two states is 
\begin{eqnarray}
&&|\Phi_{(S,m)}|^{2}=
C_{(S,m)}^{2}
\left(\frac{1}{2r}\right)^{2S}
\left(r+y_{3}\right)^{S+m}
\left(r-y_{3}\right)^{S-m} 
\left(1-\frac{S}{r^{2}}\theta C\theta\right) , \cr
&&|\Psi_{(S,m^{\prime})}|^{2}=
C_{(S,m^{\prime})}^{2}
\left(\frac{1}{2r}\right)^{2S-1}
\left(r+y_{3}\right)^{S-1/2+m^{\prime}}
\left(r-y_{3}\right)^{S-1/2-m^{\prime}}
\frac{1}{2r^{2}} \theta C\theta, 
\end{eqnarray}
where we denote the normalization factors 
in (\ref{supergeneralhopfspinor}) by 
$C_{(S,m)}$ and $C_{(S,m^{\prime})}$. 
$\Phi$ and $\Psi$ form rings on the body and are localized 
at $y_{3}=mr/S$ and 
$y_{3}=m^{\prime}r/(S-1/2)$ ($S\neq 1/2$) 
respectively.


\section{Summary and discussions}
\label{sec:summary}
\hspace{0.4cm}
In this paper, we have considered the motion of a charged particle 
moving on a supersphere in the presence of a supermonopole. 
The supermonopole was constructed by the supersymmetric first Hopf map. 
This system is a supersymmetric generalization 
of the quantum Hall system on 
a bosonic two-sphere. 
We obtained a relationship between 
the commutative coordinates and the noncommutative 
guiding center coordinates. 
It was shown that 
they were identified in the lowest Landau level. 
The guiding center coordinates form the 
algebra of the fuzzy supersphere. 
We also obtained two kinds of ground state wavefunctions 
from the Hopf spinor. 
They have the same energy 
and are related by the supersymmetry. 

We would like to comment on a relationship to 
the noncommutativity of D-branes. 
The fact that coordinates of a charged particle 
is described by noncommutative guiding center coordinates 
is related to 
the two descriptions of D-branes (which is simply explained in 
the second paragraph in the introduction). 
See \cite{hep-th/9910053} for the discussion of spherical D$2$-branes. 
The number of D$0$-branes is expressed by the size of matrix 
(or noncommutative coordinate) in the 
D$0$-brane's description. 
On the other hand, it is expressed by the first 
Chern number of a magnetic monopole from 
the viewpoint of a D$2$-brane. These two quantities 
are given by $2S+1$ and $2S$ respectively for the bosonic spherical 
system 
reviewed in section \ref{sec:reviewofbosonictwospheresystem}. 
The agreement of these two quantities can be seen in 
the limit of large $S$, which implies the fact that 
the two systems provide the same descriptions in this limit. 
We expect that such a comparison 
can be done in the supersymmetric system though 
an interpretation of D-brane is not clear. 
We evaluated the Chern number in (\ref{superchernnumber}) 
and found that it was given by the contribution only from the body 
space. Therefore it gave the same value as the bosonic 
monopole. 
It can be compared with the body of the superalgebra. 
The $osp(1|2)$ superalgebra contains 
the $su(2)$ subalgebra whose representation 
is decomposed into the spin $S$ and $S-1/2$ representations. 
It is natural to regard 
the spin $S$ representation of the $su(2)$ subalgebra 
as the body of the $osp(1|2)$ superalgebra. 
Thus the comparison in the supersymmetric system 
resulted in that in the bosonic system. 

A new element compared to the bosonic system is 
the existence of the supersymmetry. 
We used 
a supersymmetric extension of the first Hopf map 
based on the supergroup $OSp(1|2)$. 
The supermonopole constructed from the map 
showed the supersymmetric structure (\ref{susyfieldstrength}). 
Since the noncommutativity is expected to stem from 
the monopole field strength, 
such a structure leads to a supersymmetric 
noncommutativity. 
We also obtained the ground state wavefunctions which 
were given by the Hopf spinor. 
Their supersymmetric structure is explicitly shown in 
(\ref{superspinSwavefnctionsusytr}). 
The supersymmetric structure is naturally included 
due to the use of the supergroup $OSp(1|2)$. 

We conclude this section with a future problem. 
The extension to higher dimensional systems 
remains as an interesting problem. 
A bosonic higher dimensional quantum Hall system was 
first constructed in 
\cite{cond-mat/0110572}. 
Further generalizations have been discussed in 
\cite{hep-th/0203264,cond-mat/0306045,cond-mat/0306351}. 
We have investigated 
a relationship between such bosonic higher dimensional systems 
and noncommutative geometry in \cite{hasebekimura,hep-th/0402044}. 
Since the appearance of noncommutative geometry 
in higher dimensional systems is different from 
two dimensional systems, 
it is important to study 
how noncommutative superspaces arise 
in higher dimensional supersymmetric systems.


\vspace{1.0cm}
\begin{center}
{\bf Acknowledgments}
\end{center}
\hspace{0.4cm}
The authors would like to thank S. Iso and H. Umetsu 
for helpful discussions. 
This work was supported in part by JSPS Research Fellowships 
for Young Scientists. 


\renewcommand{\theequation}{\Alph{section}.\arabic{equation}}
\appendix

\section{Notation}
\setcounter{equation}{0} 
\label{sec:notation}
\hspace{0.4cm}
In this section, we summarize some notations which are related to 
supermatrix (superalgebra). 
Let $X$ be a supermatrix: 
\begin{equation}
X=\left( \begin{array}{cc} 
 A& B\\ 
 C&D 
  \end{array}\right),
\end{equation}
where $A$ and $D$ are even elements, while $B$ and 
$C$ are odd (Grassmann) elements. We define 
the superadjoint operation $\ddagger$ as 
\begin{equation}
X^{\ddagger}=\left( \begin{array}{cc} 
 A ^{\dagger}& C^{\dagger} \\ 
 -B ^{\dagger} & D^{\dagger} 
  \end{array}\right),
\end{equation}
where $\dagger$ means the usual adjoint operation. 
The superadjoint operation on the $osp(1|2)$ generators is, 
therefore, given by 
\begin{equation}
l_{i}^{\ddagger}=l_{i}, \hspace{0.4cm}
v_{\alpha}^{\ddagger}=C_{\alpha\beta}v_{\beta}. 
\end{equation}

We next define superstar $\ast$ which act on Grassmann numbers 
as 
\begin{equation}
(\theta_{i}\theta_{j})^{\ast}
=\theta_{i}^{\ast}\theta_{j}^{\ast}, 
\hspace{0.4cm}
\theta_{i}^{\ast\ast}=-\theta_{i}. 
\end{equation}
The action on bosonic numbers is the usual complex conjugation. 
It acts on Grassmann coordinates of $S^{2,2}$ as 
\begin{eqnarray}
\theta_{\alpha}^{\ast}=C_{\alpha\beta}\theta_{\beta}, \hspace{0.4cm}
(\theta C\theta)^{\ast}=\theta C\theta. 
\end{eqnarray}


\section{$osp(2|2)$ algebra in supermonopole background}
\label{sec:osp(2|2)algebra}
\setcounter{equation}{0} 
\hspace{0.4cm}
In this section, we investigate the $osp(2|2)$ algebra 
in the supermonopole background. 
It plays an important role in constructing 
the Hamiltonian (discussed in the latter part of this section) 
or gauge field theories \cite{hep-th/9507074,hep-th/0311005}. 

The generators of the $osp(2|2)$ algebra are given by 
those of the $osp(1|2)$ algebra as 
\begin{eqnarray}
&&J^{\gamma}=
-\frac{2}{r}(\theta C)_{\beta}J_{\beta} \cr
&& \hspace{0.55cm}
=\frac{1}{r}\hbar x_{i}
(\theta \sigma_{i})_{\beta}
\frac{\partial}{\partial\theta_{\beta}}, \cr
&&J_{\alpha}^{d}=\frac{1}{r}(\sigma_{i})_{\beta\alpha}
\left(\theta_{\beta}J_{i}+x_{i}J_{\beta}
\right) \cr
&&\hspace{0.55cm}=
-\frac{r}{2}\hbar\left(1+\frac{\theta C\theta}{2r^{2}}\right)
C_{\alpha\beta}\frac{\partial}{\partial\theta_{\beta}}
-\frac{1}{2r}\hbar(\theta\sigma_{i}\sigma_{j})_{\alpha}x_{i}
\frac{\partial}{\partial x_{j}} . 
\label{osp12generatorsfromosp22}
\end{eqnarray}
A relationship between the $osp(1|2)$ algebra and 
the $osp(2|2)$ algebra is further explained in 
the appendix \ref{sec:repofosp22}. 
They satisfy the following algebra, 
\begin{eqnarray}
&&[J^{\gamma},J_{\alpha}]=\hbar J_{\alpha}^{d}, \hspace{0.4cm} 
[J^{\gamma},J_{\alpha}^{d}]=\hbar J_{\alpha}, \hspace{0.4cm} 
[J^{\gamma},J_{i}]=0,\cr
&&[J_{i},J_{\alpha}^{d}]=
\frac{1}{2}\hbar(\sigma_{i})_{\beta\alpha}J_{\beta}^{d}, \hspace{0.4cm} 
\{J^{d}_{\alpha},J^{d}_{\beta}\}
=-\frac{1}{2}\hbar(C\sigma_{i})_{\alpha\beta}J_{i}, \hspace{0.4cm} 
\{J_{\alpha},J_{\beta}^{d}\}
=-\frac{1}{4}\hbar C_{\alpha\beta}J^{\gamma}. 
\end{eqnarray}

We next study how the above 
commutation relations are deformed in the 
presence of the supermonopole background (\ref{chargeSmonopole}). 
The contribution of the gauge field is added by making 
the replacements (\ref{gaugefieldreplacement}) 
in (\ref{osp12generatorsfromosp22}):
\begin{eqnarray}
&&\Lambda^{\gamma}=-\frac{2}{r}
C_{\alpha\beta}\theta_{\alpha} \Lambda_{\beta}, \cr
&&\Lambda_{\alpha}^d=\frac{1}{r}(\sigma_i)_{\beta\alpha}
(\theta_{\beta}\Lambda_{i}+x_i \Lambda_{\beta}). 
\end{eqnarray}
They transform under the $osp(1,2)$ transformation 
as 
\begin{eqnarray}
&&[L_i,\Lambda^{\gamma}]=0,~~~
[L_{\alpha},\Lambda^{\gamma}]=-\hbar\Lambda_{\alpha}^d, \cr
&&[L_i,\Lambda^d_{\alpha}]=
\frac{1}{2}\hbar(\sigma_i)_{\beta\alpha}\Lambda^d_{\beta},~~~ 
\{L_{\alpha},\Lambda^d_{\beta}\}
=-\frac{1}{4}\hbar C_{\alpha\beta}\Lambda^{\gamma}, 
\end{eqnarray}
where $L_{i}$ and $L_{\alpha}$ are given in 
(\ref{LandLambda}). 
Since we have added the contribution of the gauge field, 
$\Lambda_{\alpha}^{d}$ and $\Lambda^{\gamma}$ no longer satisfy 
the $osp(2|2)$ algebra. 
We find that 
total angular momentum operators which satisfy the 
$osp(2|2)$ algebra are 
\begin{eqnarray}
&&L^{\gamma}=-\frac{2}{r}C_{\alpha\beta}\theta_{\alpha} 
L_{\beta}-4\hbar S , \cr
&&L^d_{\alpha}=\frac{1}{r}(\sigma_i)_{\beta\alpha}
(\theta_{\beta}L_{i}+x_i L_{\beta}),  
\end{eqnarray}
which are related to $\Lambda^{\gamma}$ and 
$\Lambda_{\alpha}^{d}$ as 
\begin{eqnarray}
&&L^{\gamma}=\Lambda^{\gamma}-\frac{2\hbar S}{r^{2}}
(\theta C \theta)-4\hbar S, \cr
{\rule[-2mm]{0mm}{8mm}\ } 
&&L_{\alpha}^d=\Lambda_{\alpha}^d+\frac{\hbar S}{r^{2}}
(\sigma_i)_{\beta\alpha}(\theta_{\beta}x_i+x_i \theta_{\beta}). 
\end{eqnarray}
They transform under the $osp(1|2)$ transformation 
as 
\begin{eqnarray}
&&[L_i,L^{\gamma}]=0,
~~~[L_{\alpha},L^{\gamma}]=-\hbar L_{\alpha}^d, \cr
&&[L_i,L^d_{\alpha}]=\frac{1}{2}\hbar(\sigma_i)_{\beta\alpha}L^d_{\beta},
~~~ 
\{L_{\alpha},L^d_{\beta}\}=-\frac{1}{4}\hbar C_{\alpha\beta}L^{\gamma}. 
\end{eqnarray}
An $OSp(1,2)$ invariant quantity can be constructed 
from $L^{\gamma}$ and $L_{\alpha}^d$ as 
\begin{equation}
C_{\alpha\beta}L_{\alpha}^d L_{\beta}^d 
+\frac{1}{4} (L^{\gamma})^2. 
\end{equation}
By replacing $L^{\gamma}$ and $L_{\alpha}^d$ with 
$\Lambda^{\gamma}$ and $\Lambda_{\alpha}^d$ respectively, 
we can construct another $OSp(1,2)$ invariant quantity.  
These are related as 
\begin{equation}
C_{\alpha\beta}L_{\alpha}^d L_{\beta}^d +\frac{1}{4} (L^{\gamma})^2
=C_{\alpha\beta}\Lambda_{\alpha}^d \Lambda_{\beta}^d 
+\frac{1}{4} (\Lambda^{\gamma})^2+4 S^2. 
\end{equation}


\vspace{0.4cm}

In the last part of this section, we comment on the Hamiltonian. 
The Hamiltonian we analyzed in the section \ref{sec:fuzzysphereinlll} 
actually does not provide a complete form of 
the kinetic term of a particle moving 
on $S^{2,2}$. 
It was written only by the $osp(1|2)$ generators. 
A complete Hamiltonian is constructed by using 
both of the $osp(1|2)$ and $osp(2|2)$ generators. 

Let us consider the following two $osp(1|2)$ invariant 
quantities: 
\begin{eqnarray}
&&H_{1}=\frac{1}{2mr^{2}}\left(J_{i}J_{i}
+C_{\alpha\beta}J_{\alpha}J_{\beta}\right), \cr
&&H_{2}=\frac{1}{2mr^{2}}\left(
C_{\alpha\beta}J_{\alpha}^{d}J_{\beta}^{d}
+\frac{1}{4}J^{\gamma 2}
\right), 
\end{eqnarray}
where $H_{1}$ is the Hamiltonian used 
in the section \ref{sec:fuzzysphereinlll}. 
Considering the following replacements 
\begin{equation}
-i\frac{\partial}{\partial x_{i}}\rightarrow p_{i}, 
\hspace{0.4cm}
-i\frac{\partial}{\partial \theta_{\alpha}}\rightarrow 
(Cp)_{\alpha}, 
\end{equation}
we rewrite the above Hamiltonians as 
\begin{eqnarray}
&&H_{1}=\frac{1}{2mr^{2}}\left[
\left(x_{i}x_{i}+\frac{1}{4}(\theta C\theta) \right)
p_{j}p_{j}
+\frac{1}{4}\left(x_{i}x_{i}-\frac{1}{2}(\theta C\theta)\right)
(pC p) +\frac{3}{2}i\epsilon_{ijk}x_{j}p_{k}
(\theta\sigma_{i}Cp)\right], \cr
&&H_{2}=\frac{1}{2mr^{2}}\left[
-\frac{1}{4}(\theta C\theta) 
p_{j}p_{j}
-\frac{1}{8}\left(2r^{2}+\theta C\theta\right)
(pCp) +\frac{1}{2}i\epsilon_{ijk}x_{j}p_{k}
(\theta\sigma_{i}Cp)\right].
\end{eqnarray}
We have used 
the following two relations, 
\begin{eqnarray}
&&x_{i}x_{i}+C_{\alpha\beta}\theta_{\alpha}\theta_{\beta}
=r^{2}, \cr 
&&x_{i}p_{i}+C_{\alpha\beta}\theta_{\alpha}p_{\beta}
=0.
\end{eqnarray}
Therefore the following linear combination 
realizes the kinetic term of a particle on $S^{2,2}$: 
\begin{eqnarray}
H_{1}-3H_{2}=\frac{1}{2mr^{2}}\left(p_{i}p_{i}
+C_{\alpha\beta}p_{\alpha}p_{\beta}\right). 
\end{eqnarray}
We see that this combination can be the Hamiltonian without 
the gauge field.


\section{The representation theory of 
$OSp(1|2)$ and $OSp(2|2)$}
\setcounter{equation}{0} 
\label{sec:repofosp22}
\hspace{0.4cm}
In this section, we review the representation theory of 
$OSp(1|2)$ and $OSp(2|2)$. 
See \cite{PaisRittenberg,ScheunertNahmRittenberg,Marcu} 
for references. 

We denote the $osp(1|2)$ generators 
by $\{l_i,v_{\alpha}\}$, where $i=1,2,3$ and $\alpha=1,2$. 
The bosonic part forms the $su(2)$ algebra. 
The $osp(1,2)$ algebra is given by 
\begin{equation}
[l_i,l_j]=i\epsilon_{ijk}l_k, \hspace{0.4cm}
[l_i,v_{\alpha}]=\frac{1}{2}(\sigma_i)_{\beta\alpha}v_{\beta},
\hspace{0.4cm}
\{v_{\alpha},v_{\beta}\}=\frac{1}{2}(C\sigma_i)_{\alpha\beta}l_i. 
\end{equation}
The irreducible representation of $OSp(1|2)$ is characterized by 
an integer or half-integer $l$ which is called superspin. 
This representation is decomposed into the spin $l$ and 
$(l-1/2)$ representations of $SU(2)$. 
The dimension is 
$(2l+1)+2l=4l+1$. 
The quadratic Casimir is given by 
\begin{equation}
l_{i}l_{i}+C_{\alpha\beta}v_{\alpha}v_{\beta}
=l\left(l+\frac{1}{2}\right). 
\end{equation}

\vspace{0.6cm}

We next consider the $OSp(2|2)$ group. 
Let $\{l_i,v_{\alpha},d_{\alpha},\gamma\}$ 
($i=1,2,3$, $\alpha=1,2$) 
be a basis of the $osp(2|2)$ algebra 
forming 
\begin{eqnarray}
&&[l_i,l_j]=i\epsilon_{ijk}l_k,
~~[l_i,v_{\alpha}]=\frac{1}{2}(\sigma_i)_{\beta\alpha}v_{\beta},
~~\{v_{\alpha},v_{\beta}\}=\frac{1}{2}(C\sigma_i)_{\alpha\beta}l_i,\cr
&&[\gamma,l_{\alpha}]=d_{\alpha},
~~~~~~[\gamma,d_{\alpha}]=v_{\alpha},
~~~~~[\gamma,l_i]=0,\cr
&&[l_i,d_{\alpha}]=\frac{1}{2}(\sigma_i)_{\beta\alpha}d_{\beta},
~~~~\{d_{\alpha},d_{\beta}\}=-\frac{1}{2}(C\sigma_i)_{\alpha\beta}l_i,\cr 
&&\{v_{\alpha},d_{\beta}\}=-\frac{1}{4}C_{\alpha\beta}\gamma.
\end{eqnarray}
The bosonic part of the $osp(2|2)$ algebra 
forms $su(2)\oplus u(1)$ subalgebra 
whose generators are $\{l_i, \gamma\}$. 
The $osp(2|2)$ algebra 
contains the $osp(1|2)$ subalgebra $\{l_i,v_{\alpha}\}$ 
and  has the automorphism such as 
\begin{equation}
\{l_i,v_{\alpha},d_{\alpha},\gamma\} 
\rightarrow \{l_i,v_{\alpha},-d_{\alpha},-\gamma\}.
\label{autoOSP22}
\end{equation}
The $osp(2|2)$ algebra has two Casimir invariants:
\begin{eqnarray}
&&C_2=(l_i^2+C_{\alpha\beta}v_{\alpha}v_{\beta})
-(C_{\alpha\beta}d_{\alpha}d_{\beta}+\frac{1}{4}\gamma^2),\cr
&&C_3=\frac{1}{2}\gamma C_2+\frac{1}{2} \gamma 
C_{\alpha\beta}(v_{\alpha}v_{\beta}-d_{\alpha}d_{\beta}) 
+\frac{1}{3}(\sigma_i C)_{\alpha\beta}
(-v_{\alpha}l_i d_{\beta}+d_{\alpha} l_i v_{\beta})\cr
&& \hspace{1.0cm}+\frac{1}{6}(\sigma_i C)_{\alpha\beta}
(-v_{\alpha}d_{\beta}+d_{\alpha}v_{\beta})l_i . 
\label{Casimirosp22}
\end{eqnarray}

\vspace{0.4cm}

We summarize the irreducible representations of $osp(2|2)$. 
They are classified into two categories.
One is called typical representation 
and the other non-typical representation. 
The typical representation is reducible with respect to $osp(1|2)$ 
and is not 
specified by the two Casimirs of $osp(2|2)$ 
since both of them vanish. 
On the other hand, 
the non-typical representation is irreducible with respect to $osp(1|2)$ 
and is specified by the two Casimirs of $osp(2|2)$.
Any representations of $osp(2|2)$ are reducible 
with respect to $u(1)\oplus su(2)$ and are 
constructed by the direct sum of 
irreducible representations of $u(1)\oplus su(2)$. 
We label the representations by $(g; j,j_3)$:  
\begin{eqnarray}
&l_i^2|g; j,j_3\rangle=j(j+1)|g; j,j_3\rangle,\label{OSP22l2}\cr
&l_3|g; j,j_3\rangle=j_3|g; j,j_3\rangle,\label{OSP22l3}\cr
&\gamma|g; j,j_3\rangle=2g|g; j,j_3\rangle,\label{OSP22g}
\end{eqnarray}
where $j=0,\frac{1}{2},1,\cdots$, 
$j_3=-j,-j+1,\cdots,j$ and 
$g$ takes an arbitrary complex number. 

The irreducible representations of $osp(2|2)$  
are classified into the following four cases. 

\vspace{0.2cm}

\noindent 
\underline{$(g;0)$}
\vspace{0.2cm}

\noindent 
This is a trivial one-dimensional representation. 

\vspace{0.2cm}

\noindent 
\underline{$(j;j)$}
\vspace{0.2cm}

\noindent
This is a $4j+1$ dimensional representation and is decomposed into 
\begin{equation}
|j;j,j_3\rangle \oplus |j+1/2;j-1/2,j_3\rangle. 
\end{equation}
It is a non-typical representation since it is irreducible 
with respect to $osp(1|2)$. 
$\{d_{\alpha},\gamma\}$ are constructed from 
$\{l_i,l_{\alpha}\}$ as 
\begin{eqnarray}
&&\gamma=\frac{1}{j+\frac{1}{4}}
\biggl(C_{\alpha\beta}v_{\alpha}v_{\beta}
+2j\biggl(j+\frac{1}{2}\biggr)\biggr),\label{gamma} \cr
&&d_{\alpha}=[\gamma,l_{\alpha}]=
-\frac{1}{2\left(j+\frac{1}{4}\right)}
(\sigma_i)_{\beta\alpha}(v_{\beta}l_i+l_iv_{\beta})\label{dalpha}.
\end{eqnarray}
Since both of the Casimirs (\ref{Casimirosp22})
vanish 
\begin{equation}
C_2=C_3=0, 
\label{Cas0}
\end{equation}
they do not specify the 
irreducible representation.

\vspace{0.2cm}

\noindent 
\underline{$(-j;j)$}
\vspace{0.2cm}

\noindent 
This case is related to the $(j;j)$ representation by the 
automorphism of $osp(2,2)$ (\ref{autoOSP22}) since 
the sign of $g=j$ changes when we change the sign of 
$\gamma$. 
It is also a non-typical 
and $4j+1$ dimensional representation: 
\begin{equation}
|-j;j,j_3\rangle \oplus 
|-j-1/2;j-1/2,j_3\rangle .
\end{equation}
This representation is what was used in the appendix 
\ref{sec:osp(2|2)algebra}.

\vspace{0.2cm}

\noindent 
\underline{$(g;j)$ ($2g\neq j$)}
\vspace{0.2cm}

\noindent 
This representation is $8j$ ($j\neq 0$) dimensional one, which is 
decomposed into 
\begin{equation}
|g;j,j_3\rangle \oplus 
|g+1/2;j-1/2,j_3\rangle \oplus 
|g-1/2;j-1/2,j_3\rangle \oplus
|g,j-1,j_3\rangle .
\end{equation}
The first two representations 
form the superspin $j$ irreducible representation 
of $osp(1,2)$, 
while the last two do the superspin 
$j-\frac{1}{2}$ irreducible representation. 
Thus this representation is a  typical one: 
\begin{equation}
(g;j)_{osp(2,2)}\rightarrow (j)_{osp(1,2)}\oplus (j-1/2)_{osp(1,2)}.
\end{equation}  
The Casimirs are given by
\begin{equation}
C_2=j^2-g^2,~~~~C_3=g(j^2-g^2). 
\end{equation}


\section{Derivation of (\ref{commutationrelationofosp12lambda})}
\setcounter{equation}{0} 
\label{sec:derivationof}
\hspace{0.4cm}
In this section, we show the detailed derivation of 
the equation (\ref{commutationrelationofosp12lambda}). 
Commutation relations of $\Lambda_{i}$ and $\Lambda_{\alpha}$ 
are calculated as 
\begin{eqnarray}
&&[\Lambda_{i},\Lambda_{j}]=i\hbar\epsilon_{ijk}
\left(\Lambda_{k}+\tilde{\Lambda}_{k}^{(1)} \right), \cr 
{\rule[-2mm]{0mm}{5mm}\ } 
&&[\Lambda_{i},\Lambda_{\alpha}]=
\frac{1}{2}\hbar(\sigma_{i})_{\beta\alpha}
\left(\Lambda_{\beta}+\tilde{\Lambda}_{\beta}\right), \cr
{\rule[-2mm]{0mm}{5mm}\ } 
&&\{\Lambda_{\alpha},\Lambda_{\beta}\}=
\frac{1}{2}\hbar(C\sigma_{i})_{\alpha\beta}
\left(\Lambda_{i}+\tilde{\Lambda}_{i}^{(2)}\right). 
\end{eqnarray}
where 
\begin{eqnarray}
&&\tilde{\Lambda}_{i}^{(1)}=\frac{1}{2}\epsilon_{jkl}x_{k}x_{i}
F_{jl}
+\frac{i}{2}x_{j}(\theta\sigma_{j})_{\alpha}F_{i\alpha}
-\frac{i}{2}x_{i}(\theta\sigma_{j})_{\alpha}F_{j\alpha}
+\frac{1}{8}\epsilon_{ijk}(\theta \sigma_{j}F\sigma_{k}^{T}\theta), 
\cr 
&&\tilde{\Lambda}_{\alpha}=
-\frac{1}{3}\epsilon_{ijk}(\theta\sigma_{l}\sigma_{i})_{\alpha}
x_{j}F_{kl}
+\frac{1}{3}\epsilon_{ijk}(C\sigma_{i}\sigma_{l})_{\alpha\beta}
x_{j}x_{l}F_{k\beta}, \cr
&&\hspace{2.0cm}
+\frac{i}{6}x_{i}(\theta FC\sigma_{i})_{\alpha}
-\frac{i}{3}\theta_{\alpha}x_{i}tr(FC\sigma_{i}) \cr
&&\tilde{\Lambda}_{i}^{(2)}
=-\frac{1}{4}(\theta C\theta)\epsilon_{ijk}F_{jk}
-\frac{i}{2}(\sigma_{k}\sigma_{i}\sigma_{j})_{\alpha\beta}
\theta_{\alpha}x_{j}F_{\beta k}
+\frac{i}{4}tr(C\sigma_{i}F)x_{j}x_{j}
-\frac{i}{2}tr(C\sigma_{j}F)x_{j}x_{i}. 
\hspace{0.6cm}
\end{eqnarray}
These are greatly simplified after we substitute the values 
of the field strength (\ref{chargeSmonopole}):
\begin{eqnarray}
\tilde{\Lambda}_{i}^{(1)}=\tilde{\Lambda}_{i}^{(2)}
=-\frac{eg}{r}x_{i}, \hspace{0.4cm}
\tilde{\Lambda}_{\alpha}
=-\frac{eg}{r}\theta_{\alpha}. 
\end{eqnarray}


%
%
%



\end{document}